\documentclass[aps,prd,preprintnumbers,nofootinbib,onecolumn]{revtex4}%
\usepackage{color} 
\usepackage{amsmath,braket}
\usepackage{amssymb}
\usepackage{bm}
\usepackage{graphicx} 
\newcommand{\bea}{\begin{eqnarray}}
\newcommand{\eea}{\end{eqnarray}}
\newcommand{\be}{\begin{equation}}
\newcommand{\ee}{\end{equation}}

 \makeatletter  
\def\alt{\mathrel{\mathpalette\gl@align<}}
\def\agt{\mathrel{\mathpalette\gl@align>}}
\def\gl@align#1#2{ \lower.6ex\vbox{\baselineskip\z@skip\lineskip\z@
\ialign{ $\m@th#1\hfil##\hfil$\crcr#2\crcr\sim\crcr }} } \makeatother

\begin{document}

%

\title{
Unruh radiation produced by a uniformly accelerating charged particle in thermal random motions
}
\vspace{1cm}

%
\author{
Naritaka Oshita${}^{1,2}$, Kazuhiro Yamamoto${}^{3,4}$, and Sen Zhang${}^5$
}
 \vspace{.5cm}

\affiliation{
$^{1}$Research Center for the Early Universe (RESCEU), Graduate School of Science,
The University of Tokyo, Bunkyo-ku, Tokyo 113-0033, Japan\\
$^{2}$Department of Physics, Graduate School of Science,
The University of Tokyo, Bunkyo-ku, Tokyo 113-0033, Japan \\
$^{3}$Department of Physical Science, Graduate School of Science,
Hiroshima University, Higashi-Hiroshima 739-8526, Japan\\
$^{4}$Hiroshima Astrophysical Science Center, Hiroshima University,
         Higashi-Hiroshima 739-8526, Japan \\
$^{5}$Okayama Institute for Quantum Physics,
Kyoyama 1-9-1, Kita-ku, Okayama 700-0015, Japan
}

\vspace{2cm} 
\begin{abstract}
In this study, we investigate the signature of the Unruh effect in quantum radiation from an accelerated charged
particle interacting with vacuum fluctuations. Because a charged particle in uniformly
accelerated motion exhibits thermal random motion around the classical trajectory because of the
Unruh effect, its quantum radiation might be termed Unruh radiation.
We show that the energy flux of the quantum radiation is negative and that its amplitude is
smaller than the classical Larmor radiation by a factor of $a/m$, where $a$ is the acceleration
and $m$ is the mass of the particle.
The total radiation flux of the classical Larmor radiation and the quantum radiation is positive;
therefore, the quantum radiation appears to suppress the total radiation.
Interestingly, the results are consistent with the prediction for the quantum correction to
classical Larmor radiation, which were obtained using a different approach.
\end{abstract}
\maketitle

\def\Omega{\sigma}
\section{Introduction}\label{intro}
Hawking radiation refers to the theoretically predicted black hole emission of thermal radiation
with a temperature of $T_{\text{BH}} = 1/8 \pi G M$, where $M$ is the mass of the black hole and $G$
is the gravitational constant \cite{HawkingRadiation}.
The detection of Hawking radiation is extremely difficult because the Hawking
temperature of a black hole with solar mass $M_{\odot} \sim 2 \times 10^{30}$ [kg]
is of the order of $T_{\text{BH}} \sim10^{-7}$ [K], which is much smaller than the temperature
of the cosmic microwave background $T_{\text{CMB}} \simeq 2.7$ [K].
The Unruh effect refers to the prediction that a uniformly accelerated observer
sees the Minkowski vacuum as a thermal bath with a temperature of $T_{\text{U}} = a/ 2 \pi$,
where $a$ is the acceleration of the observer \cite{Unruh}.

The Unruh effect is related to the Hawking radiation via the principle of equivalence.
Therefore, any experimental verification of the Unruh effect will be useful
as an analogy of the Hawking radiation.
P. Chen and T. Tajima pointed out the possibility of testing the Unruh effect using an
extremely intense laser \cite{ChenTajima}.
An electron accelerated by the laser will see the Minkowski vacuum as a thermal bath; then,
radiation might be produced via the Unruh effect, which is referred to as Unruh radiation.
Using a laser with an intensity of the order of $10^{25}$ [W/cm$^2$] \cite{ELI}, the Unruh
temperature reaches $T_{\text{U}} \simeq 10^{3}$ [K], which is much higher than room temperature.
Thus, Unruh radiation is interesting as evidence of the Unruh effect.

Possible signatures of Unruh radiation have been theoretically investigated by several authors.
The authors in Refs.\citenum{ChenTajima,Schutzhold,Schutzhold2} investigated the Unruh radiation
produced by scattering between an accelerated electron and the Rindler particle of
the thermal bath originating from the Unruh effect.
However, the importance of quantum interference in predicting the Unruh radiation has been
pointed out \cite{Raine,Raval,LinHu,IYZ13,IYZ,OYZ15}; we take this interference into account in our theoretical
investigation of the signature of the Unruh radiation \cite{OYZ16}.

The paper is organized as follows. In Sec.~II, we review our results in Ref.\citenum{OYZ16}.
Adopting a model consisting of a charged particle and an electromagnetic field,
we calculate the energy flux while taking the quantum interference term into account.
In Sec.~III, we show that the results in Sec.~II are consistent with the first-order
quantum correction to the Larmor radiation, which was previously
obtained using a different approach in Refs.\citenum{HW,NSY,YN,KNY,NY}.

\section{Quantum radiation produced by an accelerated charged particle}
We adopt a model consisting of a point particle (with a mass $m$ and a charge $e$)
and an electromagnetic field $A^{\mu} (x)$, which are coupled to each other.
The action of the model is given by $S_{\text{tot}} = S_{\text{P}} + S_{\text{EM}} + S_{\text{int}}$ with
$S_{\text{P}} = -m \int d\tau \sqrt{\eta_{\mu \nu} \dot{z}^{\mu} \dot{z}^{\nu}}$, $S_{\text{EM}}
= - \int d^4 x F^{\mu \nu} F_{\mu \nu}/4$, and $S_{\text{int}} (z, A^{\mu}) = -e \int d\tau \int d^4x \delta^4_{\text{D}} (x-z(\tau)) \dot{z}^{\mu}(\tau) A_{\mu} (x)$, where
$F_{\mu \nu} = \partial_{\mu} A_{\nu} -\partial_{\nu} A_{\mu}$ is the field strength and $\eta_{\mu \nu}$ is the metric
of the Minkowski spacetime, for which we follow the convention $(+---)$.
The equations of motion are derived from the action as follows:
\begin{eqnarray}
&&m \ddot{z}_{\mu} = e(\partial_{\mu} A_{\nu} - \partial_{\nu} A_{\mu}) \dot{z}^{\nu} +f_{\mu}(A_{\text{BG}}),
\label{032302}\\
&&\partial_{\mu} \partial^{\mu} A^{\nu} = e \int d \tau \dot{z}^{\nu} (\tau) \delta^{4}_{\text{D}} (x -z(\tau)),
\label{032301}
\end{eqnarray}
\begin{eqnarray}
\text{with} \ \partial_{\mu} A^{\mu} = 0 \ \text{(gauge condition)},
\end{eqnarray}
where we introduce the external force $f_{\mu}$ that accelerates the particle with acceleration $a$
using a uniform electromagnetic field $A_{\text{BG}}$. The solution of Eq.~(\ref{032301}) is given by
\begin{eqnarray}
&&A_{\mu} (x) =A_{\text{BG}\mu} (x) + A_{\text{h}} {}_{\mu} (x) + e\int d\tau G_{\text{R}} (x,z(\tau)) \dot{z}_{\mu}(\tau),
\label{032303}
\end{eqnarray}
where $A_{\text{h}}^{\mu}$ satisfies $ \partial_{\nu} \partial^{\nu} A^{\mu}_{\text{h}} = 0$, and
$G_{\text{R}}(x,y)$ represents the retarded Green's function that satisfies
$\partial_{\mu} \partial^{\mu} G_{\text{R}}(x,y) = \delta^4_{\text{D}} (x-y)$.
Note that the background part of the electromagnetic field $A_{\text{BG}} (x)$
does not contribute to the radiation energy flux, but instead, simply uniformly accelerates the particle.
We introduce the inhomogeneous solution of the electromagnetic field $A_{\text{inh}}^{\mu} (x)$
as follows
\begin{eqnarray}
A_{\text{inh}}^\mu (x) \equiv e\int d\tau G_{\text{R}} (x,z(\tau)) \dot{z}^{\mu}(\tau).
\end{eqnarray}
Decomposing the trajectory of the particle $z_{\mu}=z_\mu(\tau)$ into two parts
\begin{eqnarray}
&& z^{\mu} = \bar{z}^{\mu} + \delta z^{\mu}
\label{032401}\\
&&\bar{z}^{\mu} = (a^{-1} \sinh{(a \tau)}, a^{-1} \cosh{(a\tau)},0,0),
\label{032501}
\end{eqnarray}
and substituting (\ref{032303}) into (\ref{032302}),
we obtain the linearized equation of (\ref{032302}),
\begin{eqnarray}
m \delta \ddot{z}^i (\tau) = \frac{e^2}{6 \pi} (\delta \dddot{z}^i -a^2 \delta \dot{z}^i)
+ e(\eta^{i \nu} \dot{\bar{z}}^{\alpha} - \eta^{i \alpha} \dot{\bar{z}}^{\nu}) \partial_{\nu} A_{\text{h}} {}_{\alpha}(x)|_{x =z(\tau)}.
\label{mddotz}
\end{eqnarray}
In the following discussion, we omit the third-order time derivative term of the radiation reaction
force $(e^2/6 \pi)\delta \dddot{z}^i$ in (\ref{mddotz}), because
the contribution of this term to the perturbation $\delta z^i$ is of the order
of ${\mathcal O}((a/m)^2)$ \footnote{We need to impose the
condition $a \ll m$ so as to avoid the boiling vacuum that is due to the Schwinger effect. Therefore, $a/m$ is much smaller than unity.},
as mentioned in the paper by S. Zhang \cite{Zhang:2013ria}.
Moreover, this term corresponds to the effect caused by the short-distance
dynamics, whose scale is of the order of $e^2/m$, which is much smaller than the Compton
length $1/m$.

\begin{figure}[t]
  \begin{center}
   \includegraphics[width=57mm]{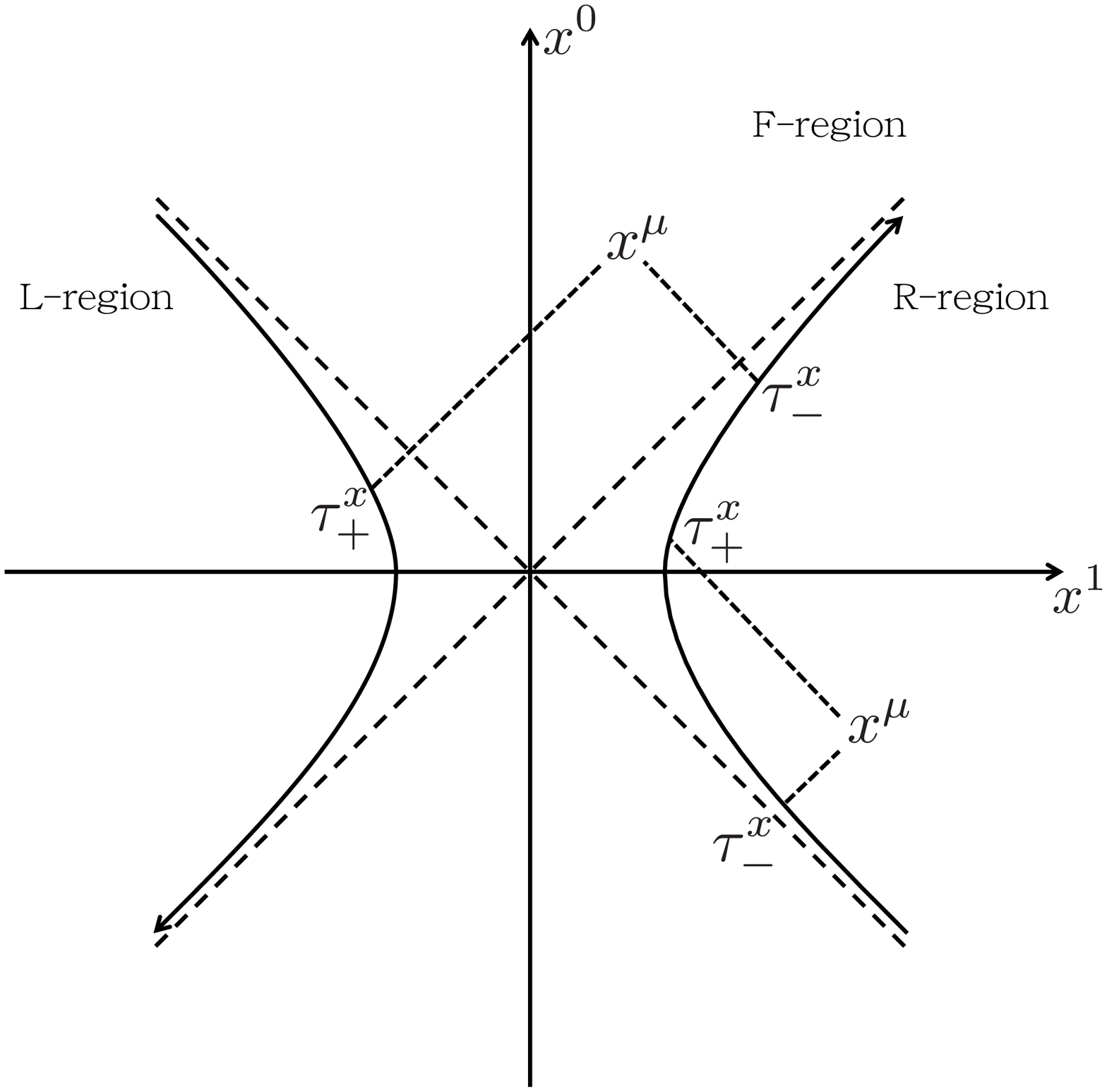}
 \includegraphics[width=63mm]{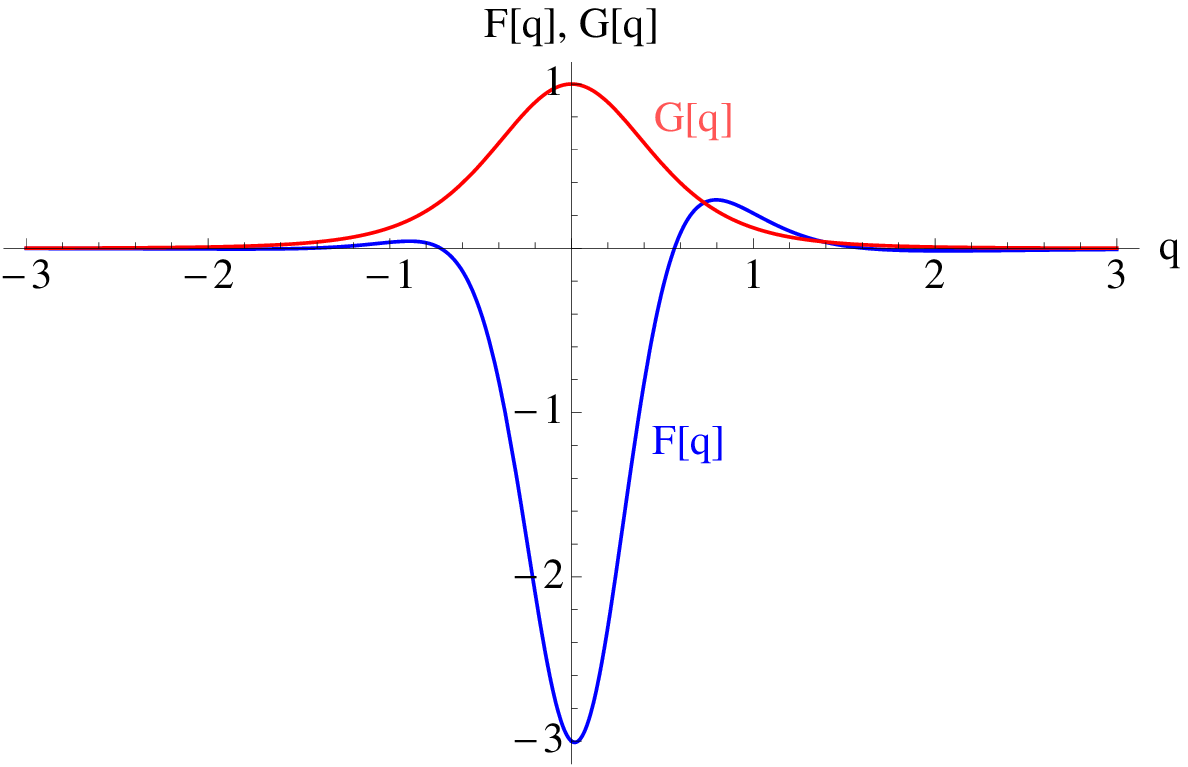}
  \end{center}
   \vspace{-0.5cm}
 \caption{
(Left) The hyperbolic curve in the R-region is the trajectory of a uniformly accelerating particle. The hyperbolic curve in the L-region is the hypothetical trajectory obtained from an analytic continuation of the true trajectory. For an observer at point $x^\mu$ in the R-region, $\tau_-^x$ is defined by the proper time of the particle's trajectory intersecting with the past light cone; $\tau_+^x$ is similarly defined by the future light cone. For an observer in the F-region, $\tau_+^x$ is the proper time of the hypothetical trajectory in the L-region intersecting with the past light cone.
(Right) Functions $G(q)$ (red line) and $F(q)$ (blue line) with
divergent terms omitted. The sign of $G(q)$ is positive; the sign of $F(q)$ is
typically negative.
\label{fig1}}
\end{figure}

To obtain the energy flux of the Unruh radiation, we first calculate the two-point correlation function,
\begin{eqnarray}
&&\langle A^{\mu}(x) A^{\nu} (y) \rangle - \langle A_{\text{h}}
^{\mu} (x) A_{\text{h}}^{\nu} (y) \rangle - A_{\text{BG}}^{\mu}(x) A_{\text{BG}}^{\nu} (y)
\nonumber \\
&&~= \langle A^{\mu}_{\text{h}} (x) A^{\nu}_{\text{inh}} (y) \rangle +
\langle A^{\mu}_{\text{inh}} (x) A^{\nu}_{\text{h}} (y) \rangle +
\langle A^{\mu}_{\text{inh}} (x) A^{\nu}_{\text{inh}} (y) \rangle,
\end{eqnarray}
where $\langle A_{\text{h}}^{\mu} (x) A_{\text{h}}^{\nu} (y) \rangle$
represents the Wightman function for vacuum fluctuations of the field and $A_{\text{BG}}^{\mu}$
does not contribute to the energy flux; therefore, we subtracted it from the
two-point function.
The inhomogeneous solution is written as
\begin{eqnarray}
&&A_{\text{inh}}^{\mu} (x) = \frac{e \dot{z}^{\mu} (\tau_-^x)}{4 \pi \rho(\tau_-^x)}
\end{eqnarray}
with
$\rho(x_-^x) \equiv \dot{z}_{\mu} (\tau_-^x) (x^{\mu} - z^{\mu} (\tau_-^x))$,
where $\tau_-^x$ satisfies the relation $(x- z(\tau_-^x))^2 = 0$ and the trajectory $z (\tau_-^x)$ is in the R-region (see
the left panel of Fig. \ref{fig1}). Using the expansion in (\ref{032401}), we obtain
$\rho(x) = \rho_0 (x) + \delta \rho(x)$,
where $\rho_0$ and $\delta \rho$ are defined as
$\rho_0 (x) \equiv \dot{\bar{z}}_{\mu} (\tau_-^x) x^{\mu}$ and $
\delta \rho (\tau_-^x) \equiv \delta \dot{z}_{\mu} (\tau_-^x) (x - \bar{z}^{\mu} (\tau_-^x))$, respectively.
Then, we obtain the formula for the inhomogeneous part $A_{\text{inh}}^{\mu} (x)$
\begin{eqnarray}
A_{\text{inh}}^{\mu} (x) = \frac{e}{4 \pi \rho_0 (x)} (\dot{\bar{z}}^{\mu} (\tau_-^x) - E^{\mu}_{(-)i}(x) \delta \dot{z}^i (\tau_-^x)),
\label{032402}
\end{eqnarray}
where we define $E^{\mu i}_{\mp} (x)$ as
$E^{\mu i}_{\mp} (x) \equiv \eta^{\mu i} - \frac{\dot{\bar{z}}^\mu (\tau_{\mp}^x) x^i}{\rho_0 (x)}$
and $\tau_\pm^x$ satisfies the relation $(x -\bar z(\tau_\pm^x))^2 = 0$ (see the left pane of Fig. \ref{fig1}).
Using (\ref{032402}), we obtain the formula (see Ref.\citenum{OYZ16} for details)
\begin{eqnarray}
&&\left[\langle A^{\mu}_{\text{h}} (x) A^{\nu}_{\text{inh}} (y) \rangle +
\langle A^{\mu}_{\text{inh}} (x) A^{\nu}_{\text{h}} (y) \rangle +
\langle A^{\mu}_{\text{inh}} (x) A^{\nu}_{\text{inh}} (y) \rangle \right]_{S} \nonumber \\
&&= \left( \frac{e}{4 \pi} \right)^2 \frac{\dot{\bar{z}}^{\alpha} (\tau_-^x)}{\rho_0 (x)} \frac{\dot{\bar{z}}^{\beta} (\tau_-^x)}{\rho_0 (x)}
+ \left[ \frac{e}{4 \pi \rho_0 (x)} \frac{e}{4 \pi \rho_0 (y)} \frac{E^{\beta}_{(-)i} (y) E^{\alpha i}_{(+)} (x)}{2m} I_2 (x,y)
\right.
\nonumber \\
&&
\left.
+ \frac{e}{4 \pi \rho_0^3 (x)} \frac{e}{4 \pi \rho_0 (y)} E^{\beta}_{(-)i} (y) x^i \eta^{\alpha A} a \epsilon_{A}^{ \ A'}x_{A'}
\frac{i}{2m} (I_1 (x,y) - I_3 (x,y)) \right] \nonumber \\
&&+ \left[ (x, \alpha) \leftrightarrow (y, \beta) \right],
\end{eqnarray}
where the functions $I_1(x,y)$, $I_2(x,y)$ and $I_3(x,y)$ are defined as
\begin{eqnarray}
&&I_1(x,y) = -\frac{i}{2 \pi \sigma} + \frac{i}{\pi} \log \left( 1 + e^{-a |\tau_-^y - \tau_+^x|} \right)
+ \frac{i}{\pi} a (\tau_-^y - \tau_+^x) \theta (\tau_-^y - \tau_+^x) + {\mathcal O}(\sigma),\\
&&I_2(x,y) = - \frac{a}{\pi} \frac{1}{e^{a (\tau_+^x)} + 1} + {\mathcal O}(\sigma),
~~~~~~~~~~~~~~~~~~~~~~~~~~~~~~~~~~~~~~~~~~~~~~~~~~~~~~~~~~ \\
&&I_3 (x,y) = - \frac{i}{2 \pi \sigma} + \frac{i}{\pi} \log{\left( 1-e^{-a |\tau^y_- - \tau^x_-|} \right)}
+ \frac{i}{\pi} a (\tau_-^y - \tau_-^x) \theta (\tau_-^y - \tau_-^x) + {\mathcal O}(\sigma),
\end{eqnarray}
for the F-region $x^0 > |x^1|$, where the dimensionless parameter $\sigma$ is defined as $\sigma \equiv e^2a/6 \pi m$.
Using the formulas of the energy-momentum tensor $T_{0 \mu} = - (A_{\alpha,0} - A_{0,\alpha}) (A^{\alpha}_{, \mu} - A_{\mu}^{,\alpha})$
and the energy flux $f = - \Sigma^3_{i = 1} T_{0i} n^i$ $(n^i \equiv x^i/\sqrt{\Sigma x^i x^i} \equiv x^i /r)$,
we obtain the energy flux expressed as $f(x) = f^C(x) + f^Q(x)$ at a large distance (i.e., $r \to \infty$).
$f^C(x)$ and $f^Q(x)$ are the classical part (the Larmor radiation) and the quantum part (the Unruh radiation),
respectively, defined by
\begin{eqnarray}
f^C = \left( \frac{e}{4 \pi} \right)^2 a^2 \frac{1}{r^2} \frac{1}{\sin^4 \theta} G (q), \ ~~~~
f^Q = \left( \frac{e}{4 \pi} \right)^2 \frac{a^3}{2 \pi m} \frac{1}{r^2} \frac{1}{\sin^4{\theta}} F(q),
\label{032601}
\end{eqnarray}
with
\begin{eqnarray}
&&G (q) = \frac{1- P^2}{(1 +q^2)^2},\\
&&F (q) = \frac{1}{(1 + q^2)^3} \left[ 6 P (2 P^2 -1) \left\{ \log{a \epsilon} - \log{(1+e^{-a|\tau_- - \tau_+|})}
-a (\tau_- -\tau_+) \theta (\tau_- - \tau_+) \right\}
+ \frac{2 P}{(a \epsilon)^2} \right. \nonumber \\
&&\left. ~~~~~~+ 2 \frac{(3 - e^{a (\tau_+ -\tau_-)}) (2- e^{a (\tau_+ - \tau_-)} (9 - e^{a (\tau_+ - \tau_-)}) )}{(1 + e^{a (\tau_+ -\tau_-)})^3} \right],
\end{eqnarray}
where we defined $\epsilon = |\tau_-^x - \tau_-^y|$ and $P = - \tanh{a (\tau_+ - \tau_-)/2}  = q/\sqrt{1+q^2}$ with $q = a (t-r-1/2a^2r)/\sin{\theta}$.
Note that $F(q)$ contains two terms that diverge in the coincidence limit $y \to x$. This divergence derives from the point particle approximation \cite{OYZ15,OYZ14}.
The approximation can be removed by taking into account the finite-size effect of the particle. Here, for simplicity, we omit the divergent terms. This procedure does not change
our conclusions, as long as the cutoff value is of the order of $a \epsilon = {\mathcal O} (1).$
The right panel of Fig. \ref{fig1} plots $G(q)$ and $F(q)$ as functions of $q$. We see that $F(q)$ is typically negative;
thus, the quantum part of the radiation flux is typically negative.
However, the total radiation flux of the classical part and the quantum part is positive;
therefore, the quantum radiation appears to suppress the total radiation.
This property is consistent with the predictions of the model based on the massless scalar field \cite{OYZ15}.

\section{Quantum correction to the Larmor radiation}
The quantum correction to the Larmor radiation has been investigated using quantum field theory for a time dependent background\cite{HW,NSY,YN,KNY,NY}.
The field theoretical approach to the quantum radiation from an accelerated charge is different
from that described in the previous section. However, the results are consistent with those presented in the previous
section. In the previous section, we adopted the unit $\hbar=1$; here, we explicitly include $\hbar$.

According the results in Refs.\citenum{HW,YN}, in the non-relativistic limit of the
velocity of a charged particle,  $|{\bf v}| \ll 1$, and the energy of the quantum Larmor
radiation is expressed as the combination of the classical Larmor radiation
$E^{(0)}$ of the zeroth order of $\hbar$, and the quantum part
$E^{(1)}$ of the first order of $\hbar$, $E=E^{(0)}+E^{(1)}$
where
\begin{eqnarray}
&&E^{(0)}={e^2\over 6\pi}\int dt (\dot v^1(t))^2,
\\
&&E^{(1)=}{e^2\hbar\over 6\pi^2m}\int dt\int dt'{\ddot v^1(t)\dot v^1(t')-\dot v^1(t)\ddot v^1(t')\over t-t'},
\label{quantumLarmor}
\end{eqnarray}
where dots indicate differentiation with respect to $t$ or $t'$.
We may use the relations $t=\bar z^0(\tau)=a^{-1}\sinh a\tau$, $\bar z^1(\tau)=a^{-1}\cosh a\tau$,
and $v^1(t)=\tanh a\tau$, for evaluating the radiation energy.
We roughly estimate the order of the radiation rate as
\begin{eqnarray}
&&\frac{dE^{(0)}}{dt} \simeq  \frac{e^2a^2}{6 \pi},
\label{032502}\\
&&\frac{dE^{(1)}}{dt} \sim -\frac{\hbar e^2a^3}{6 \pi^2 m}.
\label{032503}
\end{eqnarray}
Note that the first-order correction to the Larmor radiation (\ref{032503})
is negative and that its amplitude is of the order of $\sim \hbar e^2 a^3/m$; these two statements are
consistent with the results of (\ref{032601}).
Therefore, the quantum radiation obtained in the previous section is related
to the quantum correction to the Larmor radiation.
The quantum correction to the Larmor radiation (\ref{quantumLarmor}) is
non-local in time, which might be useful for understanding the quantum radiation
that is due to the Unruh effect.

\section*{Acknowledgements}
We would like to thank J. Yokoyama, T. Suyama, and K. Fukushima for helpful  discussions. We also thank
Prof. P. Chen for critical discussions that took place at the beginning of this study.

\end{document}